\documentclass[preprint,showpacs,showkeys,aps,prc]{revtex4-1}
\usepackage{graphicx}
\usepackage{dcolumn}
\usepackage{bm}%

\begin{document}

\title{Maximum mass of a hybrid star having a mixed phase region in the light of pulsar PSR J1614-2230}

\date{\today}

\author{Ritam Mallick}
\email{ritam.mallick5@gmail.com}
\affiliation{Nuclear Theory Group, Institute of Physics, 
        Bhubaneswar 751005, INDIA}

\begin{abstract}
Recent observation of pulsar PSR J1614-2230 with mass about 2 solar masses poses a
severe constraint on the equations of state (EOS) of matter describing stars
under extreme conditions. Neutron stars (NS) can reach the mass limits set by
PSR J1614-2230. But stars having hyperons or quark stars (QS) having boson condensates,
with softer EOS can barely reach such limits and are ruled out. QS
with pure strange matter also cannot have such high mass unless the effect of strong
coupling constant or color superconductivity are considered. In this work I
try to calculate the upper mass limit for a hybrid stars (HS) having a quark-hadron mixed phase.
The hadronic matter (having hyperons) EOS is described by relativistic mean field theory and for the 
quark phase I use the simple MIT bag model. I construct 
the intermediate mixed phase using Glendenning construction.
HS with a mixed phase cannot reach the mass limit set by PSR J1614-2230 unless I assume a 
density dependent bag constant. For such case the mixed phase region is small. 
The maximum mass of a 
mixed hybrid star obtained with such mixed phase region is $2.01 M_{\odot}$.
\end{abstract}



\maketitle

\section{Introduction}
Neutron stars (NS) are gravitationally bound, therefore the precise measurement of mass and radius 
of a NS should provide a very fine probe for the equation of state (EOS) of dense matter.
The first reasonable ideas about the composition of compact stars argued that matter are under extreme 
densities and is mainly composed of neutrons with small fractions of protons and electrons. 
Further theoretical developments and modern experimental results opened the window to other 
possibilities. The densities in the interior of neutron stars is about $3-10$ times that of the 
nuclear saturation density ($n_0 \sim 0.15\:$fm$^{-3}$). At such high densities in their interiors, the 
matter there is likely to be in a deconfined and chirally restored quark phase \cite{weber}. 

The strange matter hypothesis was first proposed by Itoh and Bodemer \cite{itoh,bodmer} and was
then improved by Witten \cite{witten}. It states that, matter at extreme density and/or temperature
are composed of almost equal number of up, down and strange quarks, called strange quark matter (SQM). 
It is also the ground state of strongly interacting matter at such extreme conditions. If this is 
true, then matter at such extreme conditions is likely to eventually convert to SQM. Such a high
density scenario is present in the interiors of a NS and therefore normal nuclear matter 
is likely to undergo a phase transition and converts to SQM. The strange matter hypothesis was
first extensively studied in the simple MIT bag model by Farhi \& Jaffe \cite{farhi}. The 
conversion process and the phase transition was further analyzed by Alcock et al. \cite{alcock}. 
The phase transition in a NS may continue up to the surface of the star or may stop inside the star.
Depending up on where this, a quark star (QS) may be of two types, a strange star (SS) or a hybrid 
star (HS). SS are stars composed only of SQM, while HS has a quark core and a hadronic exteriors.
IN the region between the quark core and hadronic outer matter, there may exist a mixed phase
region where both quarks and hadrons are present. Thus the observed pulsars are still very much
model dependent.

Recently, Demorest et al. \cite{Demorest10} found a new maximum mass limit for compact stars by 
measuring very precisely the mass of the millisecond pulsar PSR J1614-2230 to be $M=1.97 \pm 0.04\:$M$_\odot$. 
This value is much higher than any previously measured pulsar mass.
This measurement, has imposed a very severe constraints on the EOS of matter describing
the compact objects. The model of NS, without hyperons, can easily satisfy the new mass constraint.
However the presence of strangeness, either in the form of hyperons in nuclear matter or in the form
od strange quarks in quark matter, cannot easily satisfy the mass limit. So, new studies had been carried 
out to make the hyperonic EOS and quark EOS to satisfactorily explain the new mass constraint. 

Basically to satisfy the new mass limit, one has to make the EOS stiffer, which usually is softened by the 
presence of strangeness. In the hyperonic nuclear matter sector, recent studies have suggested that the 
stiffening of hyperonic EOS is possible at par with the new experimental results \cite{Bednarek2011}.
Authors also had revisited the role of vector meson-hyperon coupling \cite{Weissenborn2011b} and hyperon
potentials \cite{Weissenborn2012}, to calculate the maximum mass.

Studies prior to the discovery pulsar PSR J1614-2230 have suggested the stiffening quark matter EOS 
from the effect of strong interactions, such as one-gluon exchange or color-superconductivity 
\cite{Lugones03,Ruester04,Horvath04,Alford07,Fischer10,Kurkela10a,Kurkela10b}, which can satisfy the 
new constraint. Ozel \cite{Ozel10} and Lattimer \cite{Lattimer10} gave first studies on the implications 
of the new mass limits from PSR J1614-2230 for quark and hybrid stars in the quark bag model. Recently,
Bonanno \& Sedrakian \cite{Bonanno2012} has succeeded in obtaining massive HS. They employed  
color-superconducting quark core and very stiff hadronic EOS (like the NL3 hyperonic model or the GM3 
nuclear model).

In this work I perform an extensive study of hybrid star mass using the 
relativistic mean-field hadronic EOS together with a simple three-flavor MIT bag model quark EOS.
The model of the HS has a mixed phase intermediate region.
I would also discuss as how the understanding of more precise astrophysical 
measurements of the mass and radius of neutron stars can help 
revealing the viability of exotic quark star models. The paper is organized as follows: 
In Section II, I describe the hadronic phase and in section III I describe 
the MIT bag model. The mixed phase EOS is constructed in section IV, using the Glendenning construction. 
I present my plots and extensively describe my results for the 
EOS and the mass-radius curve in Section V. The maximum mass for the hybrid star is also calculated in 
this section. Finally in section VI, I summarize my results and draw important conclusion from them.

\section*{Hadronic phase}

At the outermost region of the star, at comparatively low densities the matter is mainly composed of 
hadrons. I use the non linear relativistic mean field (RMF) model with hyperons (TM1 parametrization) 
to describe the hadronic phase EOS. In this model the baryons interact with mean meson fields 
\cite{boguta,glen91,sghosh,sugahara,schaffner}.

The model lagrangian density includes nucleons, baryon octet ($\Lambda,\Sigma^{0,\pm},\Xi^{0,-}$) and 
leptons 
\begin{eqnarray} 
\label{baryon-lag}   
{\cal L}_H & = & \sum_{b} \bar{\psi}_{b}[\gamma_{\mu}(i\partial^{\mu}  - g_{\omega b}\omega^{\mu} - 
\frac{1}{2} g_{\rho b}\vec \tau . \vec \rho^{\mu})  \nonumber \\ 
& - & \left( m_{b} - g_{\sigma b}\sigma \right)]\psi_{b} + \frac{1}{2}({\partial_\mu \sigma \partial^\mu 
\sigma - m_{\sigma}^2 \sigma^2 } ) \nonumber \\ 
& - & \frac{1}{4} \omega_{\mu \nu}\omega^{\mu \nu}+ \frac{1}{2} m_{\omega}^2 \omega_\mu \omega^\mu - 
\frac{1}{4} \vec \rho_{\mu \nu}.\vec \rho^{\mu \nu} \nonumber \\
& + & \frac{1}{2} m_\rho^2 \vec \rho_{\mu}. \vec \rho^{\mu} -\frac{1}{3}bm_{n}(g_{\sigma}\sigma)^{3}-
\frac{1}{4}c(g_{\sigma}\sigma)^{4} +\frac{1}{4}d(\omega_{\mu}\omega^{\mu})^2 \nonumber \\
& + & \sum_{L} \bar{\psi}_{L}    [ i \gamma_{\mu}  \partial^{\mu}  - m_{L} ]\psi_{L}.
\end{eqnarray}
Leptons $L$ are non-interacting but the baryons are coupled with the scalar $\sigma$ mesons, 
the isoscalar-vector $\omega_\mu$ mesons and the isovector-vector $\rho_\mu$ mesons. 
The model constants are fitted according to the experimental results of bulk properties of nuclear
matter \cite{glen91,schaffner}.
The TM1 model explains the nuclear saturation of but cannot sufficiently models the hyperonic matter,
as it fails to reproduce the strong observed $\Lambda \Lambda$ 
attraction. This defect can be remedied by Mishustin \& Schaffner \cite{schaffner} 
by the addision of iso-scalar scalar $\sigma^*$ mesons and 
the iso-vector vector $\phi$ mesons, coupling only with the hyperons.

The detailed EOS calculation can be found in the above mentioned references  
\cite{sugahara,schaffner}, and I do not repeat them here. 

The total energy density takes the form 
\begin{eqnarray}
\varepsilon & = & \frac{1}{2} m_{\omega}^2 \omega_0^2
+ \frac{1}{2} m_{\rho}^2 \rho_0^2 + \frac{1}{2} m_{\sigma}^2 \sigma^2
+ \frac{1}{2} m_{\sigma^*}^2 \sigma^{*2} + \frac{1}{2} m_{\phi}^2 \phi_0^2
+\frac{3}{4}d\omega_0^4+ U(\sigma) \nonumber \\
& & \mbox{} + \sum_b \varepsilon_b + \sum_l \varepsilon_l  \,,
\end{eqnarray}
and the pressure can be represented as
\begin{eqnarray}
P= \sum_i \mu_i n_i - \varepsilon, 
\end{eqnarray}
where $\mu_i$ and $n_i$ is the chemical potential and number density of particle species $i$.

\section*{Quark phase}
The quark phase is modeled according to the simple MIT bag model \cite{chodos}. The
current masses of up and down quarks are extremely small, e.g.,
$5$ and $10$ MeV respectively, whereas, for strange quark the current quark mass is not 
well established, and I vary it in my calculation.
For the bag model the energy density and pressure 
can be written as 
\begin{eqnarray}
\epsilon^Q &=& \sum_{i=u,d,s} 
\frac{g_i}{2 \pi^2} \int_0^{k_F^i} dk k^2\sqrt{m_i^2 + k^2}+ 
B_G\,,\label{edec}\\ 
P^Q &=& \sum_{i=u,d,s} \frac{g_i}{6\pi^2} 
\int_0^{k_F^i} dk \frac{k^4}{\sqrt{m_i^2 + k^2}}- B_G\,, 
\label{pdec}
\end{eqnarray}
where $k_F^i=\sqrt{\mu_i^2-m_i^2}$ and $g_i$ is the Fermi momentum and degeneracy factor 
of quarks of species $i$. 
$B_G$ is the energy density difference between 
the perturbative vacuum and the true vacuum, {\rm i.e.}, the bag constant.
In this sense $B_G$ can be considered as a free parameter.

Both the hadronic and quark matter, maintains baryon number conservation, and are beta-equilibrated
and charge neutral.

\section*{Mixed phase}
With the previously described hadronic and quark EOS, Glendenning construction \cite{glen} 
gives the mixed phase regime.
The mixed phase is the baryong density range where both quarks and hadrons are present. 
In the mixed phase the hadron and the quark phases are separately charged
but the mixed phase is charge neutral as a whole. 
Thus the matter can be parametrized by the pair of electron and baryon chemical potentials
$\mu_e$ and $\mu_n$. Pressure of the two phases are made equal to maintaining mechanical equilibrium.
To satisfy the chemical and beta equilibrium conditions the chemical potential of different
particles are related to each other.
The Gibbs criterion gives the mechanical and chemical equilibrium between two phases, and is written as
\begin{equation}
P_{\rm {HP}}(\mu_e, \mu_n) =P_{\rm{QP}}(\mu_e, \mu_n) = P_{\rm {MP}}. 
\label{e:mpp}
\end{equation}
The solution of above equation gives the equilibrium chemical potentials 
of the mixed phase. As the two phases intersects one can calculate the corresponding charge densities 
of the hadronic components $\rho_c^{\rm{HP}}$ and quark components
$\rho_c^{\rm{QP}}$ separately in the mixed phase. The volume fraction occupied 
by quark matter in the mixed phase $\chi$ is given by
\begin{equation}
\chi \rho_c^{\rm{QP}} + (1 - \chi) \rho_c^{\rm{HP}} = 0.
\label{e:vol}
\end{equation}

The mixed phase energy density $\epsilon_{\rm{MP}}$ and the number density 
$n_{\rm{MP}}$ can be written as
\begin{eqnarray}
\epsilon_{\rm{MP}} &=& \chi \epsilon_{\rm{QP}} + (1 - \chi) 
\epsilon_{\rm{HP}}, \\
n_{\rm{MP}} &=& \chi n_{\rm{QP}} + (1 - \chi) 
n_{\rm{HP}}. \label{e:mpep}
\end{eqnarray}

Therefore the EOS is now a system having a charge neutral hadronic phase at lower densities, 
a charge neutral mixed phase in the intermediate region and a charge neutral quark phase at
higher densities.

\section*{Results}

\begin{figure}
\vskip 0.2in
\centering
\includegraphics[width=3.0in]{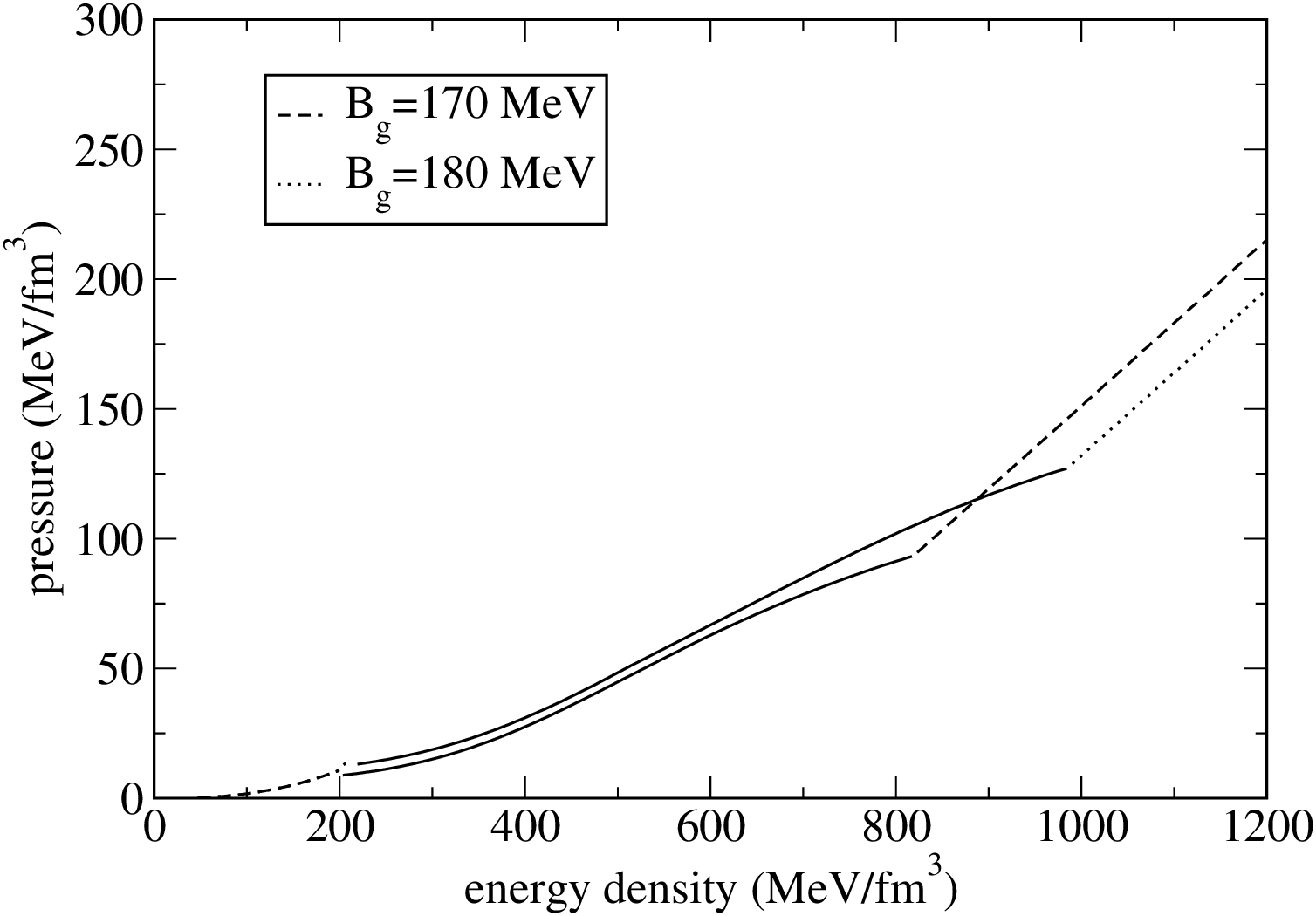}
\caption{Pressure as a function of energy density with bag pressure of $170$ and $180$MeV.}
\label{fig1}
\end{figure}

\begin{figure}
\vskip 0.2in
\centering
\includegraphics[width=3.0in]{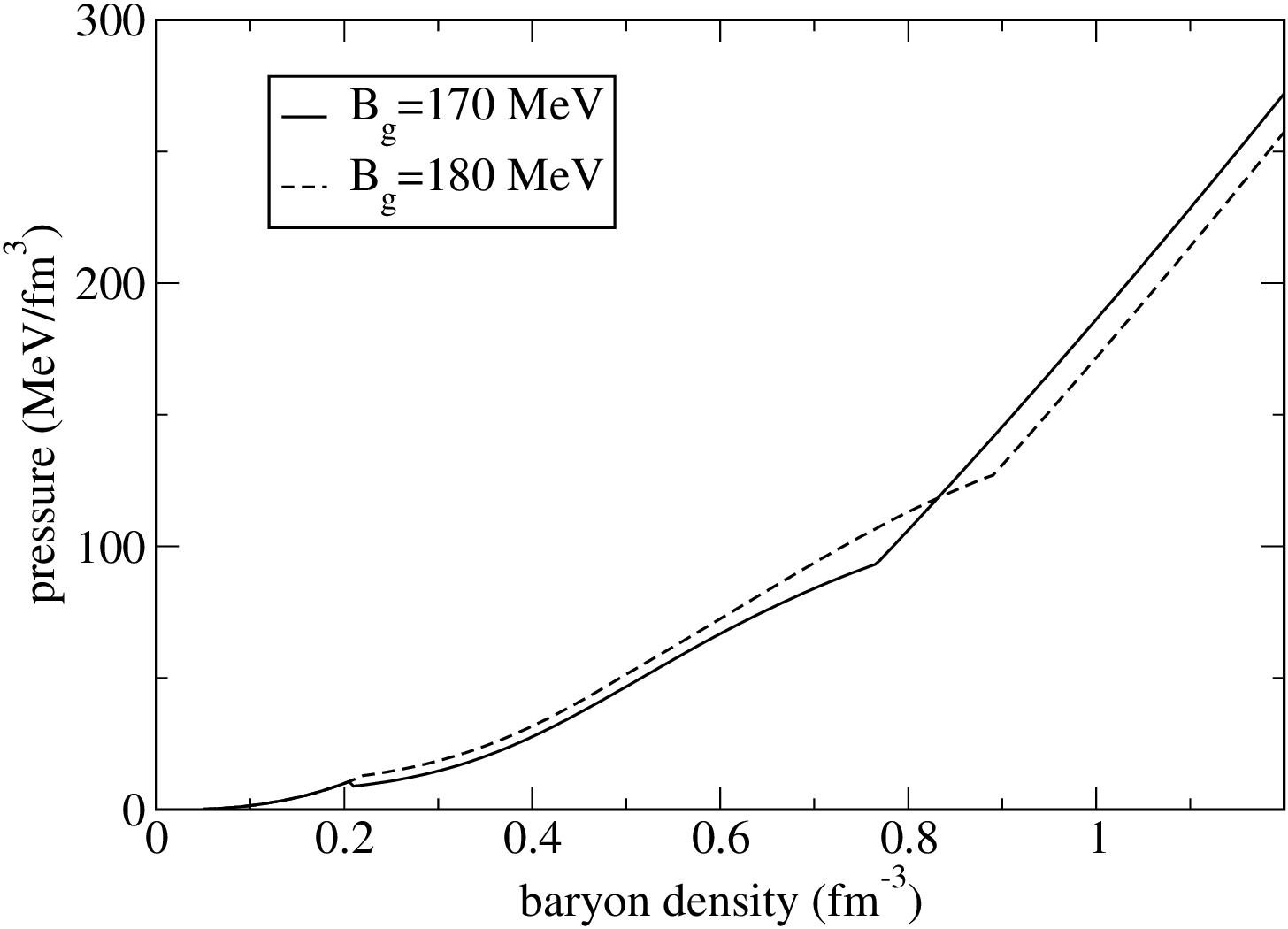}
\caption{Pressure as a function of baryon density with bag pressure of $170$ and $180$MeV.}
\label{fig2}
\end{figure}

The EOS are constructed to describe the properties of matter inside a NS, therefore
the EOS properties would also resemble the properties of a NS.  The central region of 
the star has maximum density (few times $n_0$), therefore the matter at the core is most 
likely to have a phase transition. Therefore the central region would have stable strange 
matter (or a colour superconducting matter). As the density decreases radially outwards 
some nuclear matter (nucleons) starts appearing and so in the intermediate region there is 
likely to have a mixed phase. Much further outwards I have only matter consisting of only 
nucleons. The crust consisting mainly of free electrons and nuclei, which completes the star structure.

The hadronic EOS, I assume a fixed TM1 parameter set, which satisfactorily explains the properties of 
hadronic matter at extreme condition. I can control the quark EOS by changing
the masses of the quarks and the bag constant. The masses of the light quarks 
are quite bounded and take them to be $5$MeV (u) and $10$MeV (d). The mass of s-quark is still not well
established, but expected to lie between $100-300$MeV. I would vary the mass of the 
s-quark within this bounded mass range. I would also vary the bag constant ($B_G$) to 
regulate the mixed phase region.
This parametrization of the EOS of the hadron and quark matter is responsible
for characterization of the matter in the mixed phase region.  
Using the Glendenning construction to construct the mixed phase, 
and plot curves of pressure against energy density as seen in fig \ref{fig1}.
In fig \ref{fig1} I have plotted the mixed phase EOS with bag pressures $170$MeV and $180$MeV.
Actually the relation runs as ${B_G}^{1/4}=170$MeV, but for simplicity I will denote 
${B_G}^{1/4}=170 MeV=B_g$. For this case the mass of the s-quark ($m_s$) is taken to be $150$MeV. 
With a constant bag pressure, lower bag pressure cannot generate a mixed phase region. 
I do not go above $B_g=180$MeV, as for that 
case the EOS becomes very flat, and the maximum mass of the star becomes less. In the curves, 
the lower portion is nuclear phase (dotted/dash line), the intermediate region is the mixed phase (bold 
line) and the higher region is the quark phase (dotted/dash line). Fig \ref{fig2}, shows the
pressure against baryon density for
bag constant $170$MeV  and $180$MeV. The mixed phase starts at $0.2 fm^{-3}$ and ends at
$0.76 fm^{-3}$ for bag pressure $170$MeV. For bag pressure $180$MeV the mixed phase region is in 
between $0.22 fm^{-3}$ and $0.89 fm^{-3}$. 
The curve with bag constant $170$MeV is much stiffer than the curve with bag pressure $180$MeV, 
because the bag pressure adds negatively to the matter pressure, making the effective pressure low. 
The above curves also shows that as the bag pressure 
increases the range of mixed phase region also increases. As the variation of pressure with both energy 
density and baryon density is quite similar, from now on I would only plot curve showing pressure
as function of energy density.

\begin{figure}
\vskip 0.2in
\centering
\includegraphics[width=3.0in]{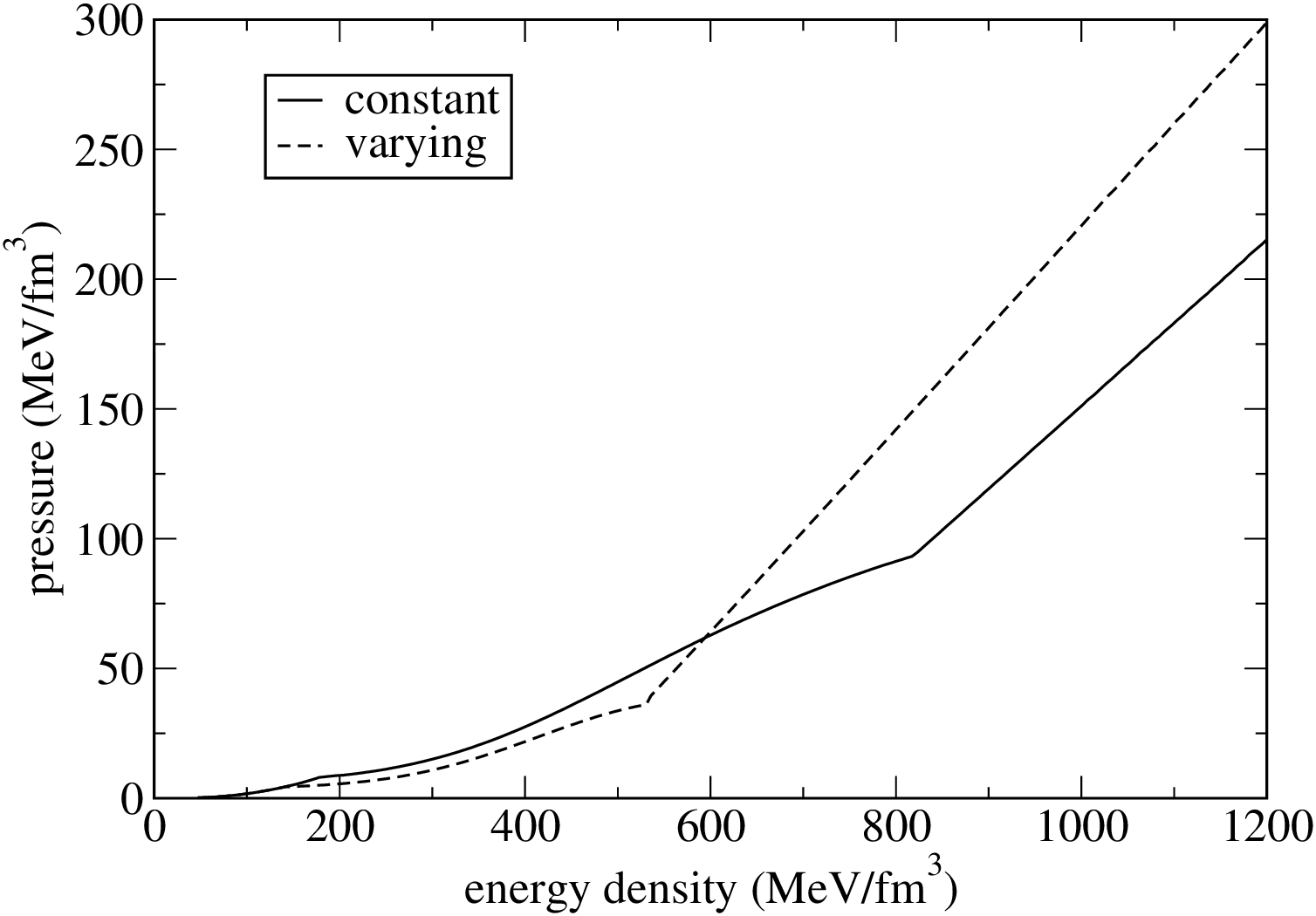}
\caption{Pressure against energy density plot with constant and varying bag pressure, having $B_g=170$MeV.}
\label{fig4}
\end{figure}

With such high bag pressure it is impossible to attain the mass limit set by PSR J1614-2230. Therefore
I have to devise some other mechanism which would give stiffer EOS, thereby increasing the maximum mass
of the HS. For that, I assume a density dependent bag constant. In the literature
there are several attempts to understand the density dependence of $B_g$ \cite{adami,blaschke}; 
however, currently the results are highly model dependent and still there is no definite picture. 
I parametrized the bag constant in such a way that it attains a value $B_\infty$, 
asymptotically at very high densities. The range of value of $B_{\infty}$ obtained from experiments can be 
found in Burgio et al. \cite{burgio}, and I assume it to be $130$MeV, the lowest value mentioned there. 
With such assumptions I then construct a Gaussian parametrization given as \cite{burgio,ritam1207}
\begin{eqnarray}
B_{gn}(n_b)  =  B_\infty  +  (B_g  -  B_\infty)  \exp  \left[  -\beta  \Big(
\frac{n_b}{n_0} \Big)^2 \right] \:. \label{bag}
\end{eqnarray}
The lowest value of $B_{gn}$, which is its value at the asymptotic high density in quark matter, is 
fixed at $130$MeV. The bag pressure quoted would be the value of the bag constant at the starting of 
the mixed phase region on the low density regime ($B_g$ in the equation). 
As the density increases the bag pressure decreases and reaches $130$MeV asymptotically, 
the decrease rate is controlled by $\beta$.

\begin{figure}
\vskip 0.2in
\centering
\includegraphics[width=3.0in]{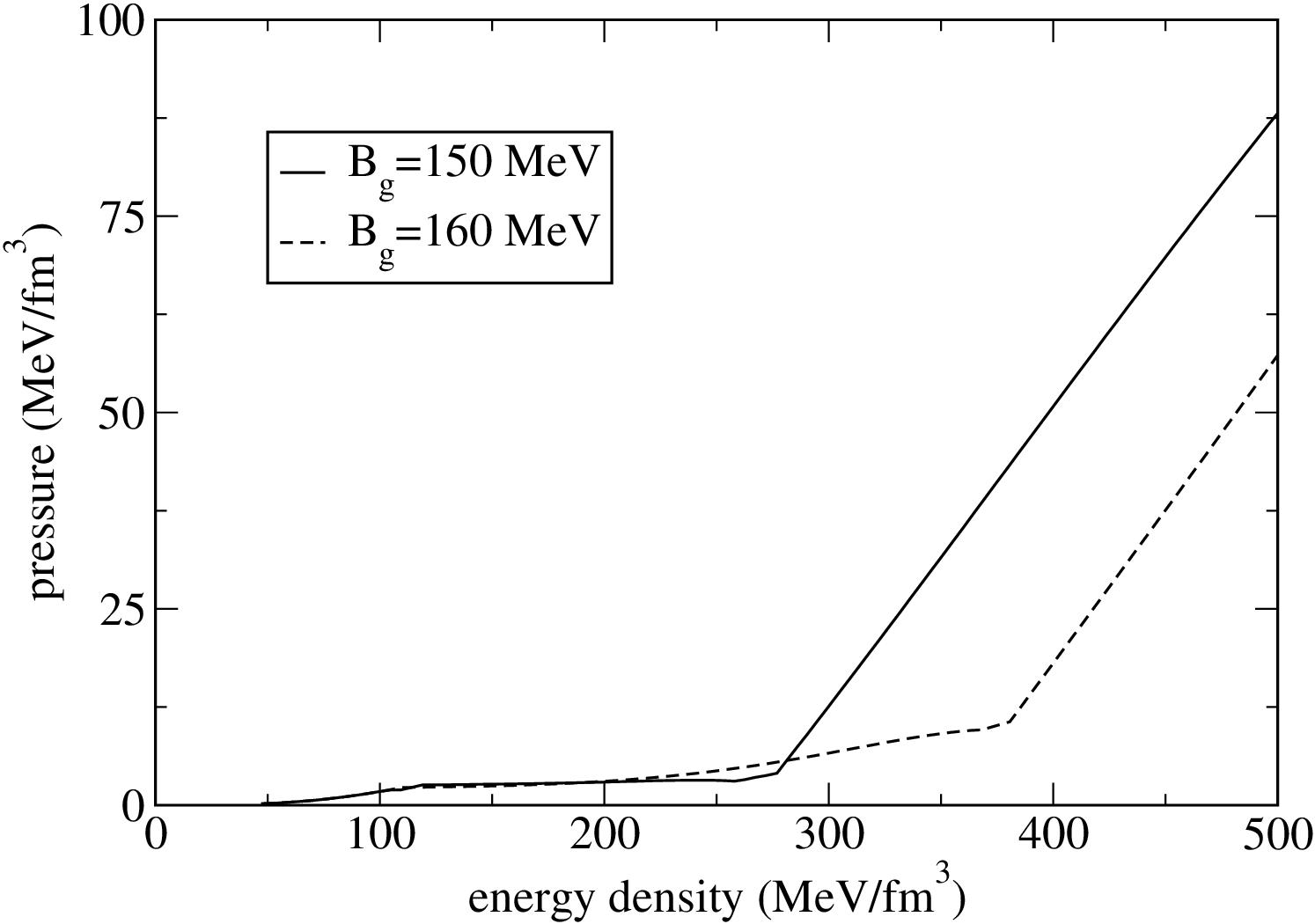}
\caption{Pressure against energy density plot with varying bag pressure, having $B_g=160$ and $150$MeV.}
\label{fig5}
\end{figure}

\begin{figure}
\vskip 0.2in
\centering
\includegraphics[width=3.0in]{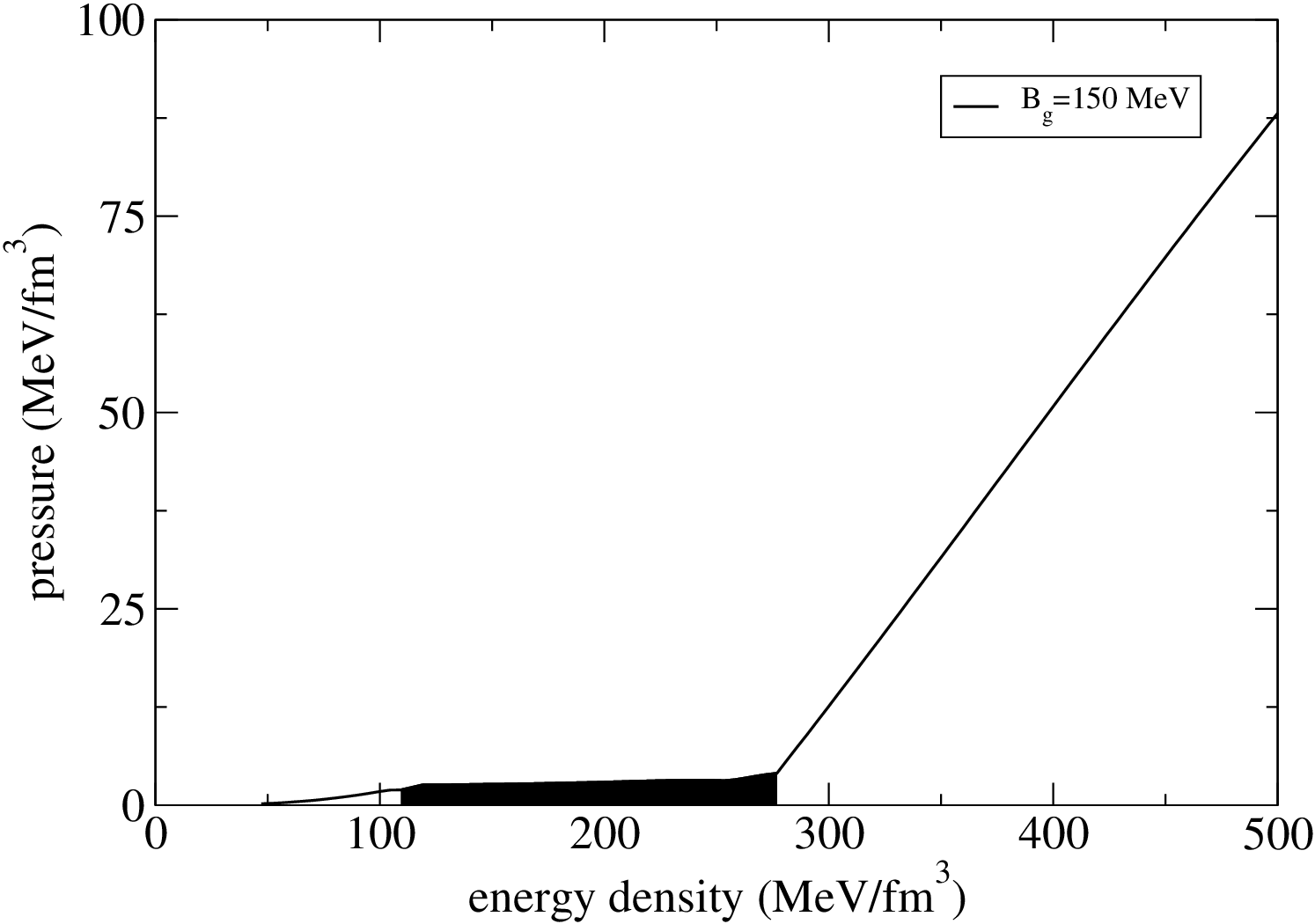}
\caption{Pressure vs energy density plot showing the explicitly the mixed phase region, for the varying bag 
pressure $B_g=150$MeV.}
\label{fig6}
\end{figure}

In fig \ref{fig4} I have plotted curves showing the difference in the slope of the curves with and
without the variation of bag pressure (for $B_g=170$MeV). For the varying bag pressure the mixed phase 
region shrinks, becomes flatter but the quark phase region becomes stiffer. 
The mixed phase region now only extends up to baryon density $0.53 fm^{-3}$. The change in the mixed phase region 
is about $\sim 30\%$. This is because, going to 
higher densities (or higher energy density towards the core) the effective matter pressure increases with the 
decrease in bag pressure (bag pressure adds negatively to the matter pressure).  
With such a density dependent bag constant I can have a significant mixed phase region with 
lower values of bag pressure.
As shown in fig \ref{fig5} I can have mixed phase region with bag pressure $B_g$, for $160$MeV and $150$MeV. 
For the  $160$MeV EOS the s-quark mass $m_s=150$MeV and for the $150$MeV curve the s-quark mass is $m_s=300$MeV. 
With bag pressure, $B_g$, $160$ and $150$MeV the mixed phase region is of considerable small. 
For bag constant $160$MeV the mixed phase region starts at density $0.15 fm^{-3}$ and ends at
$0.36 fm^{-3}$. With bag constant $150$MeV the mixed phase region starts at density $0.13 fm^{-3}$ and 
ends at $0.3 fm^{-3}$. 
In fig \ref{fig6} I have separately plotted the EOS for $B_g=150$MeV showing the mixed phase
region clearly. As it would be shown later that with only such choice of quark matter parameters I can attain
the mass limit set by PSR J1614-2230.
 
Assuming the star to be stationary and spherical, the Tolman-Oppenheimer-Volkoff (TOV) equations \cite{shapiro}
gives the solution for the pressure $P$ and the enclosed mass $m$,
\begin{widetext}
\begin{eqnarray}
  {dP(r)\over{dr}} &=& -{ G m(r) \epsilon(r) \over r^2 } \,
  {  \left[ 1 + {P(r) / \epsilon(r)} \right] 
  \left[ 1 + {4\pi r^3 P(r) / m(r)} \right] 
  \over
  1 - {2G m(r)/ r} } \:,
\\
  {dm(r) \over dr} &=& 4 \pi r^2 \epsilon(r) \:,
\end{eqnarray}
\end{widetext}
$G$ being the gravitational constant. Starting with a fixed central energy density $\epsilon(r=0) 
\equiv \epsilon_c$, I integrate radially outwards until the pressure on the surface equals the one 
corresponding to the density of iron. This gives the star's radius $R$ having gravitational mass 
\begin{equation}
M_G~ \equiv ~ m(R)  = 4\pi \int_0^Rdr~ r^2 \epsilon(r) \:. 
\end{equation}
For the NS crust, in the medium density range we add the hadronic EOS by Negele and Vautherin 
\cite{negele}, and for the outer crust we add the EOS by Feynman-Metropolis-Teller \cite{feynman} 
and Baym-Pethick-Sutherland \cite{baym}. 

\begin{figure}
\vskip 0.2in
\centering
\includegraphics[width=3.0in]{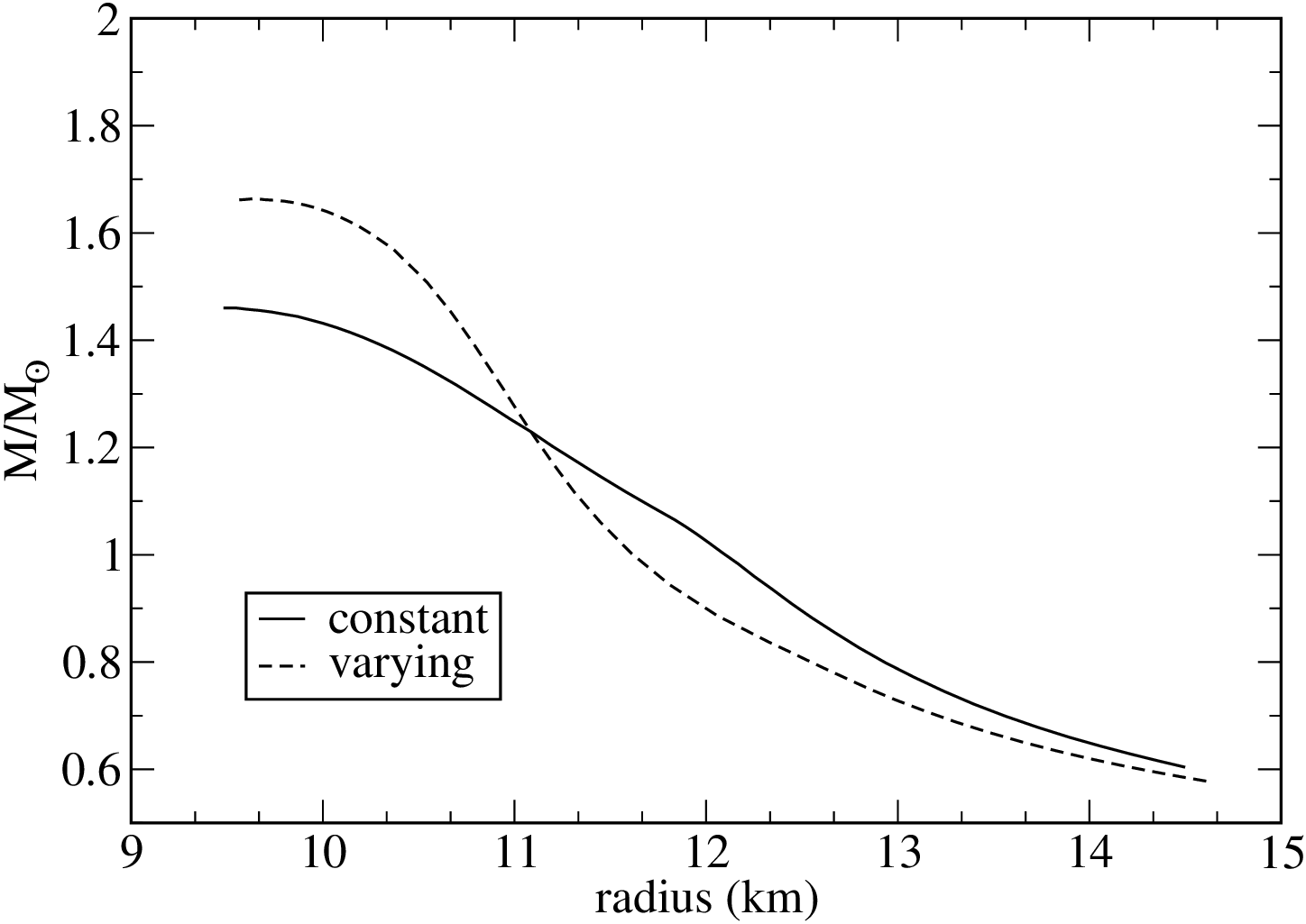}
\caption{Mass-radius curve with constant and varying bag pressure, having $B_g=170$MeV.}
\label{fig7}
\end{figure}

\begin{figure}
\vskip 0.2in
\centering
\includegraphics[width=3.0in]{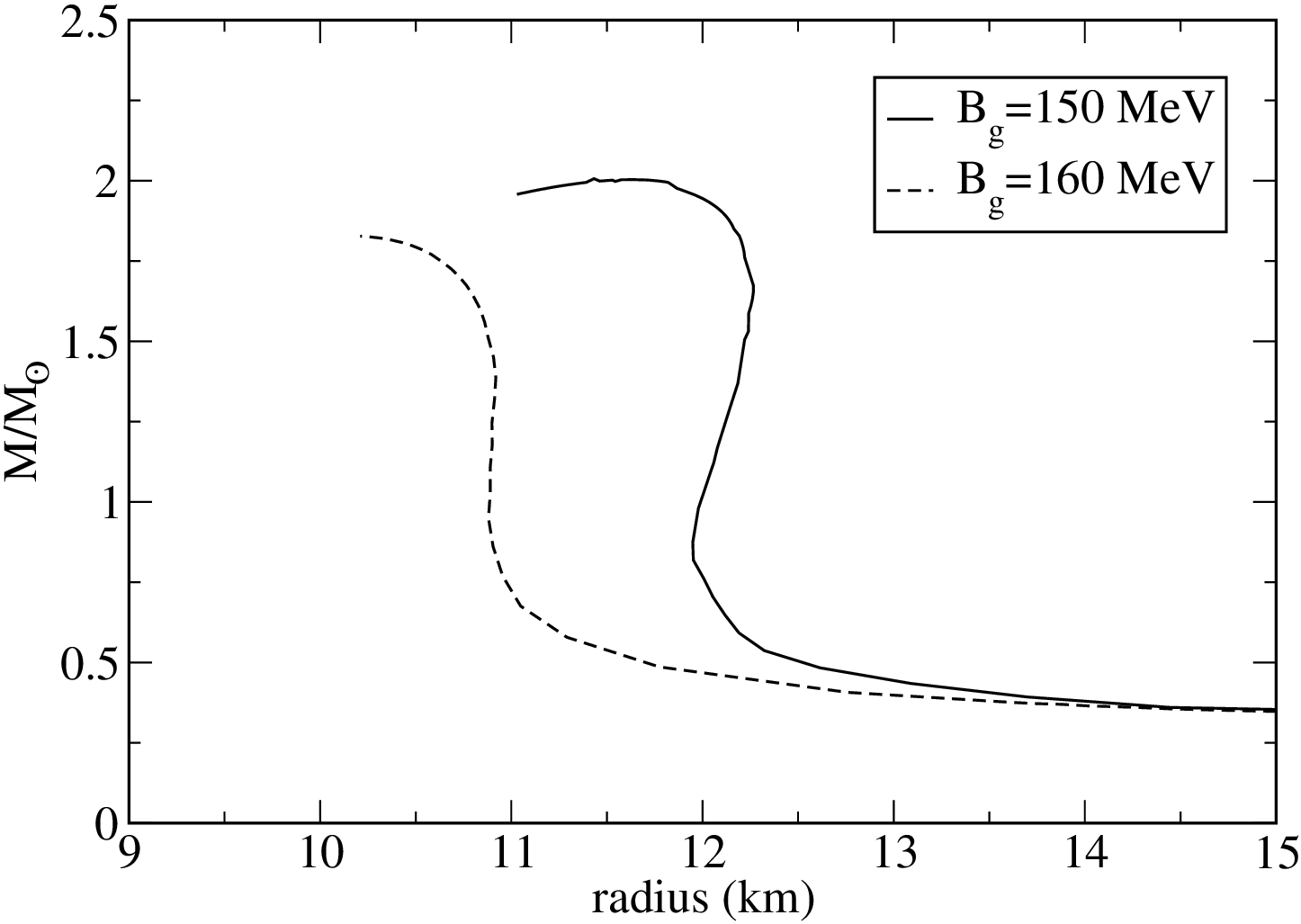}
\caption{Mass-radius curve with varying bag pressures, $B_g=160$ MeV and $150$ MeV.}
\label{fig8}
\end{figure}

Fig \ref{fig7} shows the gravitational mass $M$ (in units of solar mass $M_{\odot}$)
as a function of radius $R$, for constant and varying bag pressure $B_g=170$ MeV. 
As the bag pressure varies and decreases towards the center of the star (at higher densities) the 
curve becomes stiffer as the effective matter pressure increases (bag pressure being negative). 
I find that a flatter EOS corresponds to a flatter mass-radius curve, and therefore the maximum
mass of the star with varying bag pressure is higher than the non varying one.
With such varying bag constant I plot the mass-radius curve with $B_g=160$ MeV and $150$ MeV (fig \ref{fig8}). 
With the same qualitative aspect I find that the maximum mass of a mixed hybrid star obtained with $B_g=160$MeV 
is $1.84 M_{\odot}$. The maximum mass with $B_g=150$MeV and $m_s=300$MeV, is $2.01$ solar mass.      

The discovery of high-mass pulsar PSR J1614-2230 \cite{Demorest10} with mass of about
$1.97 M_{\odot}$, has set a stringent condition on the EOSs describing the interior of a compact star. 
They \cite{Demorest10} quote the typical values of the central density of J1614-2230, for the
allowed EOSs in the range 2$n_0$ - 5$n_0$, whereas consideration of the EOS independent analysis of 
\cite{lattimer2005} sets the upper central density limit at $10n_0$. The maximum mass of a mixed 
phase EOS star with $m_s=150$MeV is calculated to be $1.84$ solar mass. The maximum mass 
for the mixed hybrid star can be increased to $2.01$ solar mass, with $m_s=300$MeV having a varying 
bag pressure of $B_g=150$MeV. Only such choice of the quark matter parametrization can give rise to star
which would satisfy the mass set by PSR J1614-2230.
But with such choice of parameters the mixed phase region is small.
This maximum mass limit is for this hadronic and quark matter EOSs. Stiffer EOS sets (like hadronic NL3
and quark quark NJL model) for the mixed hybrid star can produce much higher maximum mass \cite{lenzi}. 
From the figure it is also clear that the maximum mass of the 
star corresponds to a radius of about $10$km. Previous calculations have shown the maximum mass of a NS have 
radius greater than $12$km, whereas the maximum mass of a SS corresponds to a radius of less than $9$km. 
Therefore it is clear from my calculation that the mixed hybrid star has radius corresponding to the 
maximum mass, quite different from the neutron and strange star. They are not as compact as strange stars 
and their radius lies between the nuclear and strange star.

\section*{Summary and Conclusion}

In this work I have studied the maximum mass of a hybrid star having a mixed phase region.
With the hadronic matter EOS having hyperons, and remaining in the simple MIT bag model I
wanted to study what parameters value could give such high masses for a HS having a mixed phase region.
The star has a dense quark core, a mixed phase intermediate region and hadronic outer region.
The hadronic and quark matter EOS is simultaneously constructed according to relativistic mean field 
approach and MIT bag
model. The mixed phase is determined in accordance with the Glendenning construction. 
All the phases are at chemical and mechanical equilibrium, and also they are 
charge neutral as a whole. With constant bag pressure $B_g$ of $170$ and $180$MeV (and $m_s=150$MeV) 
I get EOS with considerable mixed phase region but with such parametrization the maximum mass of the star 
is about $1.5$ solar mass. I therefore consider a density dependent bag pressure $B_g$, parametrized 
according to the Gaussian parametrization. The asymptotic value of the bag constant at 
high density is fixed at $130$MeV, which is its lowest value known from the experiments \cite{burgio}. 
With such varying bag 
pressure I can have a mixed phase region with $B_g=160$MeV, but still the 
mass of the star is below $1.9$ solar mass. To reach the mass limit set by PSR J1614-2230, for a mixed phase HS, 
I build the EOS with bag pressure of $B_g=150$MeV, having s-quark mass
$m_s=300$MeV. For such choice of parameters values, the mixed phase region is small. Further 
lowering of bag pressure is not possible, as then the mixed phase disappears.
The maximum mass for a mixed hybrid star with the given set of parameters is $2.01 M_{\odot}$. 
Another important results of my calculation is that the HS, with mixed phase, has radius (for the 
maximum mass) quite different from the neutron or strange star, their radius lying in between the neutron 
and strange star.

After the discovery of PSR J1614-2230, setting the mass limit to $2$ solar mass, new EOSs model
has been proposed. Weissenborn et al. \cite{Weissenborn2011a} showed that absolutely strange star can have mass 
above $2$ solar mass is the effect of strong coupling constant and color superconductivity is taken into 
account. Bednarek et al. \cite{Bednarek2011} argued that EOS with hyperons having quartic terms 
involving hidden strangeness vector meson can reach such limit. Matsuda et al. \cite{masuda} extended their
calculation to hybrid stars, having a smooth crossover from hadronic to quark matter. For the mass to reach
the maximum mass limit they showed that the crossover has to take place at low density and the quark matter
has to be strongly interacting. Using very stiff EOS sets (hadronic NL3
and quark quark NJL model) the maximum mass limit for the hybrid star can be raised much higher
as shown by Lenzi \& Lugones \cite{lenzi}. In my work, I also have shown that the maximum mass limit
can be reached by a HS with mixed phase even with simple hyperonic nuclear matter EOS and MIT bag model
quark matter EOS if I assume a relatively low density dependent bag pressure.

Observationally the NS is characterised only by signals coming to us from its surface. 
Developments has been made on them to measure accurately 
the mass of compact stars but same cannot be done for their radius. 
Reasonable measurement of the radius of a compact stars could differentiate NS, SS and HS, 
as we have seen here that different EOS of matter gives different mass-radius relationship. As it is 
clear from my calculation and also from previous calculations that by suitable tuning of the parameters or 
by invoking new terms in the EOSs calculations the mass limit set by PSR J1614-2230 can be reached. 
Therefore to have a full understanding of the matter at extreme densities we need
results not only from astrophysical observations but also from earth based 
experiments.
  

{}
\end{document}